\documentclass[sigconf,nonacm]{acmart}

\usepackage{placeins}
\AtBeginDocument{%
  }

\setcopyright{acmlicensed}
\copyrightyear{2018}
\acmYear{2018}
\acmDOI{XXXXXXX.XXXXXXX}
\acmConference[Conference acronym 'XX]{Make sure to enter the correct
  conference title from your rights confirmation email}{June 03--05,
  2018}{Woodstock, NY}
\acmISBN{978-1-4503-XXXX-X/2018/06}

\begin{document}

\title{AIDRIN 2.0: A Framework to Assess Data Readiness for AI}


\author{Kaveen Hiniduma}
\affiliation{%
  \institution{The Ohio State University}
  \country{}}
\email{hiniduma.1@osu.edu}

\author{Dylan Ryan}
\affiliation{%
  \institution{The Ohio State University}
  \country{}}
\email{ryan.1989@osu.edu}

\author{Suren Byna}
\affiliation{%
  \institution{The Ohio State University}
  \country{}}
\email{byna.1@osu.edu}

\author{Jean Luca Bez}
\affiliation{%
  \institution{Lawrence Berkeley National Laboratory}
  \country{}}
\email{jlbez@lbl.gov}

\author{Ravi Madduri}
\affiliation{%
  \institution{Argonne National Laboratory}
  \country{}}
\email{madduri@anl.gov}








\begin{abstract}
  AI Data Readiness Inspector (AIDRIN) is a framework to evaluate and improve data preparedness for AI applications. It addresses critical data readiness dimensions such as data quality, bias, fairness, and privacy. This paper details enhancements to AIDRIN by focusing on user interface improvements and integration with a privacy-preserving federated learning (PPFL) framework. By refining the UI and enabling smooth integration with decentralized AI pipelines, AIDRIN becomes more accessible and practical for users with varying technical expertise. Integrating with an existing PPFL framework ensures that data readiness and privacy are prioritized in federated learning environments. A case study involving a real-world dataset demonstrates AIDRIN’s practical value in identifying data readiness issues that impact AI model performance.
\end{abstract}

\maketitle
\vspace{-5pt}
\section{Introduction}
The AI Data Readiness Inspector (AIDRIN) \cite{10.1145/3676288.3676296} has become an essential framework for assessing and improving data preparedness for AI applications. As organizations increasingly depend on AI-driven decision-making, ensuring high-quality, AI-ready data is critical. AIDRIN addresses this need by offering a comprehensive framework for evaluating data readiness across key dimensions, including data quality, bias and fairness, and privacy. In a previous study \cite{10.1145/3722214}, we extensively examined the metrics across the pillars of data readiness. These pillars include \textit{Data Quality}, \textit{Understandability \& Usability}, \textit{Structure \& Organization}, \textit{Governance}, \textit{Impact on AI}, and \textit{Fairness}. AIDRIN categorizes the evaluation metrics within these pillars to provide users with a personalized assessment. 

This paper presents enhancements to AIDRIN by mainly focusing on user interface (UI) improvements and integration to an existing privacy-preserving federated learning (PPFL) framework. Our objective is to make AIDRIN more intuitive, efficient, and aligned with the latest research in AI data readiness. By refining the UI, we aim to increase AIDRIN's accessibility and effectiveness for both data professionals and domain experts who may not have extensive technical backgrounds in data science or AI. Additionally, integrating AIDRIN into a PPFL framework strengthens its scalability and usability by allowing smooth operation across both centralized and decentralized AI pipelines.

\begin{figure}[htbp]
    \centering
    \includegraphics[width=\linewidth]{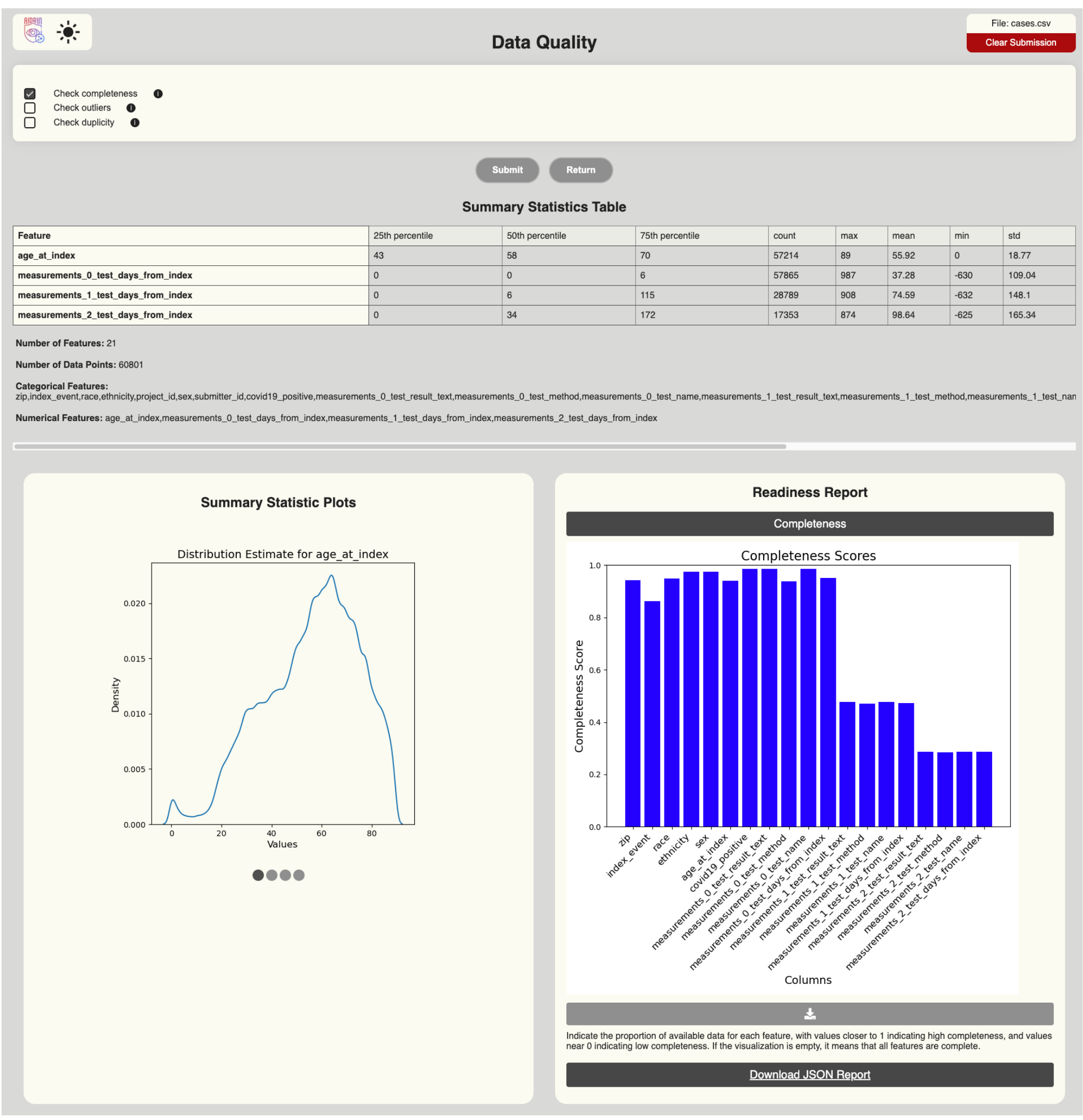}
    \caption{The figure displays the data quality page with real evaluation results from the MIDRIC cases dataset \cite{midrc_data}}
    \label{fig:ui_vis}
\end{figure}

\vspace{-5pt}
\section{UI Enhancements}
The latest version of AIDRIN introduces UI enhancements designed to improve usability and provide clearer insights. One of the key improvements is the reorganization of evaluation metrics according to the six pillars of AI data readiness, as established in our previous study \cite{10.1145/3722214}. These pillars offer a structured and comprehensive approach to assessing data readiness.

Each pillar now includes specific assessment criteria to ensure a more precise and meaningful evaluation. \textit{Data Quality} examines completeness, duplicates, and outliers to ensure the datasets meet fundamental integrity standards. \textit{Understandability \& Usability} assesses adherence to FAIR principles \cite{wilkinson2016fair} to improve data accessibility and reusability. \textit{Impact on AI} evaluates feature relevance and correlations to help users understand how their data influences model performance. \textit{Governance} analyzes data re-identification risks when quasi-identifiers are provided to address privacy concerns. Finally, \textit{Fairness} examines class imbalance, statistical parity, and representational rates of sensitive features, ensuring equitable AI outcomes.


Additionally, AIDRIN presents a table with summary statistics displaying key metrics such as mean, median, and mode, alongside visualizations of data distributions. Based on user-selected metrics, AIDRIN dynamically generates quantitative evaluations and visualizations to create a more intuitive and insightful assessment experience. These enhancements make AIDRIN a more powerful and user-friendly tool to ensure high-quality, AI-ready data.

Figure \ref{fig:ui_vis} illustrates these improvements in the UI, with the screenshot displaying the data quality page with real evaluation results generated from the MIDRC (Medical Imaging and Data Resource Center) cases dataset. The MIDRC dataset is a large, publicly available collection of de-identified medical imaging data focused on COVID-19-related chest imaging.

\section{AIDRIN Integration to APPFL Framework}

The integration of AIDRIN with APPFLx (Advanced Privacy Preserving Federated Learning as a Service) \cite{Li2023APPFLx} marks a significant advancement in the field of PPFL. APPFLx provides a user-friendly platform for conducting cross-silo PPFL experiments by enabling secure and decentralized AI model training across multiple organizations. We selected APPFLx due to its open-source nature, modular architecture, and scalability, making it a good foundation for integrating AIDRIN's data evaluation features.


AIDRIN enables edge systems to locally evaluate data and use the communication strategies in the APPFLx framework to transmit evaluation results to the server. The server then generates HTML reports using AIDRIN by combining the results from all the users involved in the FL system. These reports provide stakeholders with clear, visually organized insights that are easy to interpret and review. This approach preserves privacy by ensuring that data remains on the edge system, with only evaluation results being transmitted to the server.

By integrating AIDRIN into APPFLx \cite{li2025advances,Li2023APPFLx}, the system enables edge devices to locally evaluate data readiness before investing computational resources into training. Leveraging APPFLx’s communication framework, evaluation results are transmitted to a central server, which uses AIDRIN to generate aggregated HTML reports. These reports offer stakeholders clear, data readiness insights into data suitability across distributed clients. 

\section{Evaluations}
AIDRIN has proven to be valuable in multiple use cases since its development, particularly in Privacy-Preserving Federated Learning (PPFL) settings. One such case involved using AIDRIN to analyze data quality issues in the Flamby Heart Disease \cite{NEURIPS2022_232eee8e,Janosi1988} dataset, a real-world federated dataset collected from four different hospitals. 

Before training, AIDRIN’s analysis revealed that one client’s dataset contained only a single class, meaning all samples belonged to the same category. Additionally, the feature distribution in this client’s data was highly sparse, with one feature consisting entirely of zeros, making it impossible to compute feature correlations for that feature. These issues likely led to model bias and poor generalization, ultimately degrading overall performance.

\begin{figure}[htbp]
    \centering
    \includegraphics[width=\linewidth]{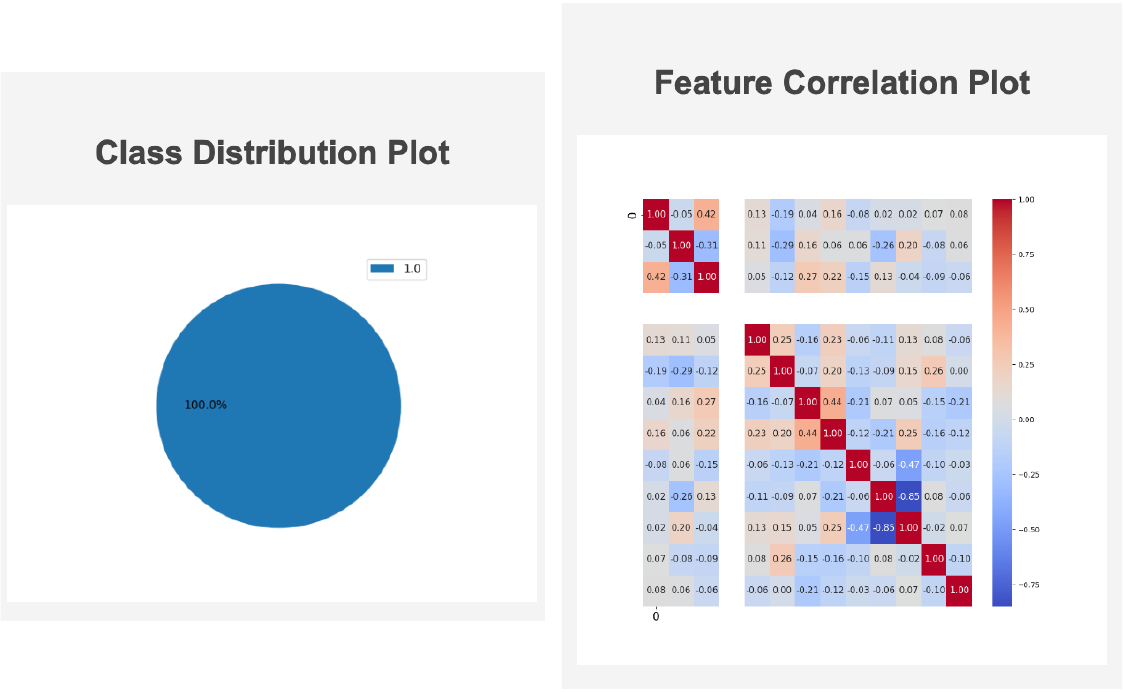}
    \caption{Visualizations from a client in the FLamby Heart Disease dataset. The left plot shows a class distribution where all samples belong to a single class, indicating complete class imbalance. The right plot presents a feature correlation matrix, where one feature could not generate meaningful feature relevances due to zero values.}
    \label{fig:results_viz}
\end{figure}

To address this, we conducted experiments where we excluded this problematic client from training, leading to a notable improvement in accuracy, from 70.6\% to 74.7\%. The initial performance drop was likely due to the imbalanced and uninformative data from the outlier client negatively influencing the global model, while removing it allowed for a more representative and effective learning process. Figure \ref{fig:results_viz} presents visualizations generated by AIDRIN by highlighting the problematic client's data characteristics, including a single-class distribution and sparse feature correlations.

\section{Conclusion}
The enhancements made to AIDRIN improve its usability and effectiveness in AI-driven environments, particularly within PPFL setting. By refining the user interface and enabling integration into decentralized AI pipelines, AIDRIN is more accessible to users of varying technical backgrounds, including those with limited expertise. The evaluation results, particularly the case study involving the Flamby Heart Disease dataset, highlight AIDRIN’s practical value in identifying critical data readiness issues that directly impact model performance. Overall, this study validates AIDRIN as a vital framework for promoting reliable, fair, and trustworthy AI development.



\FloatBarrier
\bibliographystyle{ACM-Reference-Format}
\bibliography{sample-base}

\begin{figure*}
    \centering
    \includegraphics[width=0.9\linewidth]{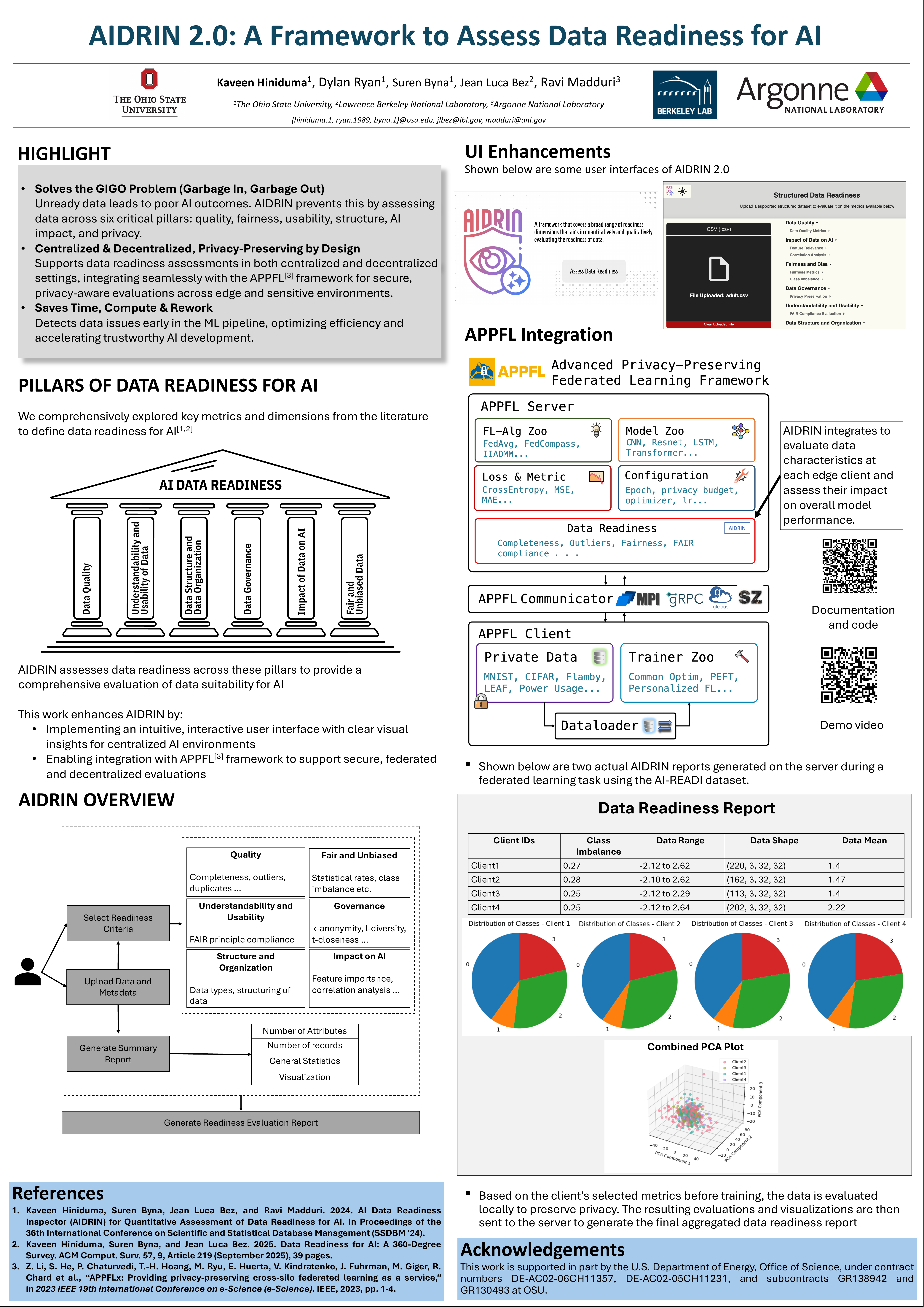}
    \caption{Accepted poster for SSDBM '25}
    \label{fig:poster}
\end{figure*}
\end{document}